\documentclass{emulateapj}

\usepackage{graphicx}
\usepackage[hypertex]{hyperref}

\def \eg {e.g.}
\def \ie {i.e.}
\def\spose#1{\hbox to 0pt{#1\hss}}
\def\ltsim{$\mathrel{\spose{\lower 3pt\hbox{$\sim$}}
        \raise 2.0pt\hbox{$<$}}$\thinspace}
\def\gtsim{$\mathrel{\spose{\lower 3pt\hbox{$\sim$}}
        \raise 2.0pt\hbox{$>$}}$\thinspace}
\newcommand{\thin }{\thinspace}

\newcommand{\mvir}{${\rm M_{vir}}$}

\newcommand{\msun }{${\rm M_{\odot}}$}

\newcommand{\tx}{${\rm T_X}$}

\newcommand{\zfe }{${\rm Z_{Fe}}$}

\newcommand{\chandra }{{\em Chandra}}

\newcommand{\minuit}{MINUIT}

\newcommand{\xspec }{{\em Xspec}}

\newcommand{\deltap}{$\delta p$}
\newcommand{\xmm }{{\em XMM}}

\newcommand{\fb}{$f_b$}

\slugcomment{Accepted for publication in the Astrophysical Journal}

\begin{document}
\title{$\chi^2$ and Poissonian data: biases even in the high-count regime and how to avoid them}
\author{\href{mailto:phumphre@uci.edu}{Philip J. Humphrey}\altaffilmark{1}, Wenhao Liu\altaffilmark{1} \& David A. Buote\altaffilmark{1}}
\altaffiltext{1}{Department of Physics and Astronomy, University of California, Irvine, 4129 Frederick Reines Hall, Irvine, CA 92697-4575}
\begin{abstract}
We demonstrate that two approximations to the $\chi^2$ statistic as popularly employed by 
observational astronomers for fitting Poisson-distributed data can 
give rise to intrinsically biased model parameter estimates, even in the high counts regime,
unless care is taken over the parameterization of the problem.
For a small number of problems, previous studies have shown that the {\em fractional} bias
introduced by  these approximations is often small when the counts are high. 
However, we show that for a broad class of problem, unless the number of data bins is far 
smaller than $\sqrt{N_c}$, where $N_c$ is the total number of counts in the dataset, the 
bias will still likely be comparable to, or even exceed, the statistical error. 
Conversely, we find that fits using \citeauthor{cash79a}'s C-statistic give comparatively unbiased 
parameter estimates when the counts are high.
Taking into account their well-known problems in the low count regime, we conclude
that these approximate $\chi^2$ methods should not routinely be used for fitting
an arbitrary, parameterized model to
Poisson-distributed data, irrespective
of the number of counts per bin, and instead the C-statistic should be adopted. We discuss several
practical aspects of using the C-statistic in modelling real data.
We illustrate the bias for two specific 
problems--- measuring the count-rate from a lightcurve and obtaining the temperature of a thermal
plasma from its X-ray spectrum measured with the \chandra\ X-ray observatory. 
In the context of X-ray astronomy, we argue the bias could give rise to systematically mis-calibrated
satellites and a $\sim$5--10\%\ shift in galaxy cluster scaling relations.
\end{abstract}
\keywords{methods: statistical--- methods: data analysis--- X-rays: galaxies: clusters--- X-rays: general}
\section{Introduction} \label{sect_introduction}
When faced with the problem of fitting a parameterized model to 
Poisson-distributed data, observational astronomers typically adopt one of two
approaches. First, the 
maximum likelihood method involves varying the model parameters
until the probability density function of the data given the model is maximal.
In practice, observers typically minimize a statistic such as C, defined by 
\citet{cash79a} which, in a slightly modified form
\citep[as implemented in the astronomical X-ray 
spectral-fitting package \xspec;][]{xspec}, can be written
\begin{eqnarray}
C& =& 2\sum_i M_i-D_i+D_i \log D_i - D_i \log M_i \label{eqn_cash}
\end{eqnarray}
where $D_i$ is the number of detected counts in the i$^{\rm th}$ data-bin, 
$M_i\equiv M_i(p_1,\ldots, p_k)$ is the model being fitted, and $p_1, \ldots, p_k$
are the model parameters.

Since the absolute value of the C-statistic cannot be directly interpreted as 
a goodness-of-fit indicator, observers typically prefer instead to minimize the 
better-known $\chi^2$ fit statistic \citep[\eg][]{lampton76a}. 
As that statistic is strictly only defined
for Gaussian-distributed data, observers generally approximate the true 
$\chi^2$ by a data-based summation of the form
\begin{equation}
\chi^2 \simeq \chi^2_{d}= \sum_i \frac{(M_i-D_i)^2}{D_i} \label{eqn_chisq_data}
\end{equation}
or
\begin{equation}
\chi^2 \simeq \chi^2_{m}= \sum_i \frac{(M_i-D_i)^2}{M_i} \label{eqn_chisq_model}
\end{equation}
where the $d$ and $m$ subscripts indicate whether the data or the model are 
used as weights.
In the literature these two weighting choices are sometimes referred to 
as ``Neyman's'' and ``Pearson's'', respectively. 
The shortcomings of these approximations are well-documented
when there are few counts per bin. 
\citet{cash79a} pointed out that deviations from Gaussianity make such 
approximations
inaccurate when the counts per bin fall below $\sim$10--20, and various
authors have quantified how the best-fitting parameters obtained
from minimizing $\chi^2_{m}$ and $\chi^2_d$ for specific models
become  biased below this limit
\citep[\eg][]{nousek89a,wheaton95a,churazov96a,leccardi07a}. 
A number of other approximations to $\chi^2$ have
been proposed to mitigate this effect
\citep[\eg][]{wheaton95a,kearns95a,churazov96a,mighell99a}. 
In contrast, at least for some problems,
fits using the C-statistic are found to be far less biased for low counts data 
\citep[\eg][]{nousek89a,churazov96a,arzner07a}, 
although not completely so \citep{leccardi07a}.

When the number of counts per bin exceeds $\sim$15--20, the deviations from 
Gaussianity become less severe. Therefore, it is common practice in observational
astronomy to assume that, in such cases, $\chi^2_d$ and $\chi^2_m$ sufficiently well
approximate the true $\chi^2$ and that the model parameters for an arbitrarily parameterized
model that minimize those statistics are relatively unbiased estimates of the true 
parameter values.
The meaning of ``relatively'' here depends on context; for most observers a 
non-negligible bias
would be acceptable provided it does not lead to the wrong scientific conclusions. 
This pragmatic approach to statistical inference is common in the observational literature,
but differs from the more rigorous methods generally preferred among statisticians. 
Nevertheless, when employing any approximation, it should be contingent upon the observer
to assess whether it could potentially lead to wrong conclusions. Unfortunately, this is 
seldom done, and approximations such as $\chi^2_d$ or $\chi^2_m$ are often used without 
comment for a given problem.

For the simple problem of measuring the count-rate of a (non-varying) source given its
lightcurve, a number of authors have assessed the accuracy of using the 
$\chi^2_d$ and $\chi^2_m$
approximations. As the count-rate becomes large, the fitted count-rate which minimizes
$\chi^2_d$ is asymptotically found to underestimate the true rate by
$\sim \tau^{-1}$ count s$^{-1}$, while similar fits using $\chi^2_m$ overestimate 
it by $\sim0.5 \tau^{-1}$~count s$^{-1}$, where $\tau$ is the duration (in seconds)
of each bin \citep{wheaton95a,jading96a,mighell99a,hauschild01a}.
This can be understood as arising from the misparameterization of the problem;
when one puts $M_i=p \tau$, where $p$ is the count-rate of the source, the
dependence of the denominator in Eqn~\ref{eqn_chisq_model}
on $p$ naturally leads to a bias. Similarly,
the dependence of the denominator in Eqn~\ref{eqn_chisq_data} on the {\em observed}
data also produces a systematic bias when minimizing $\chi^2_d$ with respect
to p \citep{wheaton95a,jading96a}.
Nonetheless, as the number of counts increases this corresponds to an increasingly 
small fractional bias. If one only requires to know the absolute count-rate to a given
fractional accuracy, therefore, the use of $\chi^2_d$ or $\chi^2_m$ may be ``good 
enough'', provided the count rate is sufficiently high.

In this paper, we point out that a more relevant quantity than the fractional
bias for assessing the usefulness of the approximations used in fitting is 
\fb, the bias {\em divided by the statistical error}. For two very 
different physical problems, obtaining the count-rate from a lightcurve and 
obtaining the temperature of a thermal plasma from its X-ray spectrum, we 
compute \fb\ for fits to realistic data which minimize $\chi^2_d$, $\chi^2_m$ and C. 
For $\chi^2_d$ and $\chi^2_m$ fits, 
we find that \fb\ can be of order unity, or even worse, even if the 
number of photons per bin far exceeds the nominal $\sim$20 counts. In contrast,
for the C-statistic fits, we find $|$\fb$|\ll$1. We explain these results
in terms of an approximate, analytical expression for \fb\ for each statistic,
and show that fits of an arbitrary, parameterized model are, in general, far less
biased when the C-statistic is employed than $\chi^2_d$ or $\chi^2_m$, unless the 
model parameterization is chosen carefully. Finally, we discuss 
the possible scientific impact of the bias, as well as the advantages and practical 
implementation of using the C-statistic instead for data-modelling. 
We stress that we are not, in this paper, attempting a formal, statistical assessment
of the validity of using $\chi^2$ methods in general to model any particular problem,
but rather we are asking whether the current {\em approximate} $\chi^2$ methods 
for Poisson-distributed data that are widely employed by observers are useful 
(in the sense $|$\fb$|\ll$1).

\section{The bias} \label{sect_bias}
In this section, we investigate two very different problems,
specifically the linear problem of obtaining the count-rate of a (non-variable) 
source from its lightcurve and the highly nonlinear problem of obtaining the 
temperature of a thermal plasma from its X-ray spectrum. We  use Monte Carlo
simulations to measure \fb\ as a function of the ``true'' parameter value, the number of 
counts in each dataset and the adopted binning.

\subsection{Lightcurve} \label{sect_lightcurve}
\begin{figure}
\centering
\includegraphics[scale=0.35]{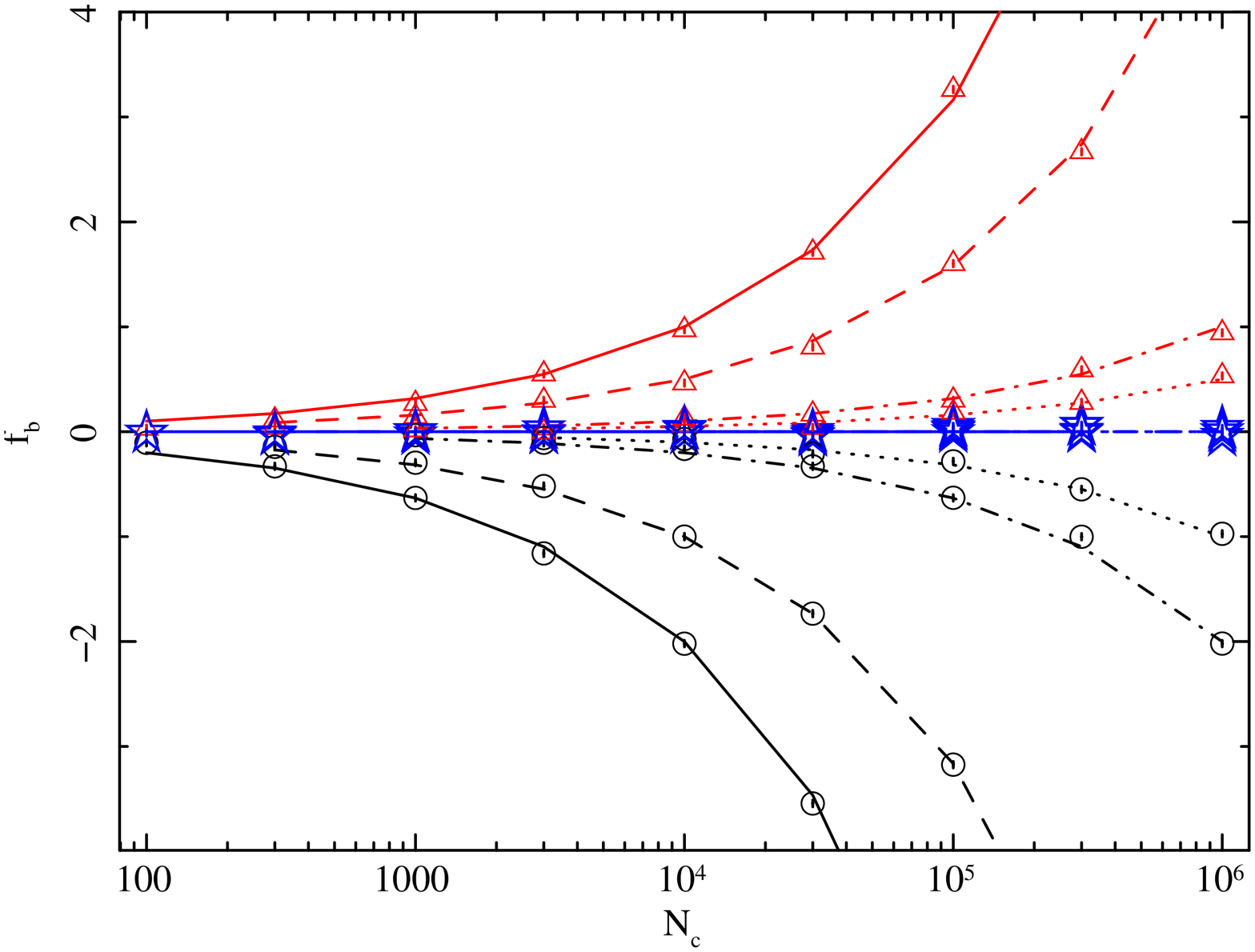}
\caption{\fb\ for the lightcurve model, as a function of total counts $N_c$ in the 
lightcurve.  Circles denote \fb\ obtained with the
$\chi^2_d$ statistic, triangles are for $\chi^2_m$ fits,
and stars are for
C-statistic minimization. Results are shown for 50, 100, 500 and 1000 counts per bin
(solid lines, dashed lines, dot-dash lines and dotted lines, respectively). The 
lines are the analytical approximations discussed in \S~\ref{sect_discussion}.} \label{fig_lightcurve}
\end{figure}

We first considered the problem $M_{i0}=p_0\tau$, where $M_{i0}$ is the true model
value in the i$^{\rm th}$ bin, $\tau$ is the binsize of the lightcurve and 
$p_0$ the true count-rate of the source. We sought an estimate of $p_0$ by fitting 
a model of the form $M_i=p\tau$ to the data. 
For each of a range of different values of $p_0$ (50, 100, 500 and 1000 count s$^{-1}$) 
and $N_c$, we simulated a set of 1000 
lightcurves with $\tau=1$\thin~s, assuming that the total
counts per bin were Poisson distributed about $p_0 \tau$. 
For each simulated lightcurve we used 
customized software built around the \minuit\
software library\footnote{http://lcgapp.cern.ch/project/cls/work-packages/mathlibs/minuit/index.html}
to obtain the value of p which minimized each statistic 
($\chi^2_d$, $\chi^2_m$ and C).
The mean and standard deviation of the best-fitting p values were measured for each ($p_0$,$N_c$) pair
and statistic choice,
allowing \fb\ to be computed. In Fig~\ref{fig_lightcurve}, we show how \fb\ varies
as a function of $N_c$, the total counts in the lightcurve, and the count-rate of the source.

As is clear from Fig~\ref{fig_lightcurve}, {\em at fixed count-rate and binsize}, the statistical
importance of the $\chi^2_d$ and $\chi^2_m$ bias is an {\em increasing} function of the 
number of counts 
in the lightcurve; indeed it rapidly becomes very large as the number of data-bins gets large.
This is simply because the absolute value of the bias is approximately constant as the 
count-rate becomes large \citep{jading96a}, whereas the statistical error is a decreasing 
function of $N_c$.
In stark contrast, for the \citeauthor{cash79a} C-statistic
fits, we find $|f_b| \ll 1$; in fact
the bias using the C-statistic is exactly zero here. This can be seen by 
substituting $M_{i}=p\tau$ into Eqn~\ref{eqn_cash} and analytically 
minimizing C, which leads to $p=\sum_i D_i/\sum_i \tau$,
the expectation of which is $p_0$.

\subsection{Thermal plasma} \label{sect_spectrum}
\begin{figure*}
\centering
\includegraphics[scale=0.35]{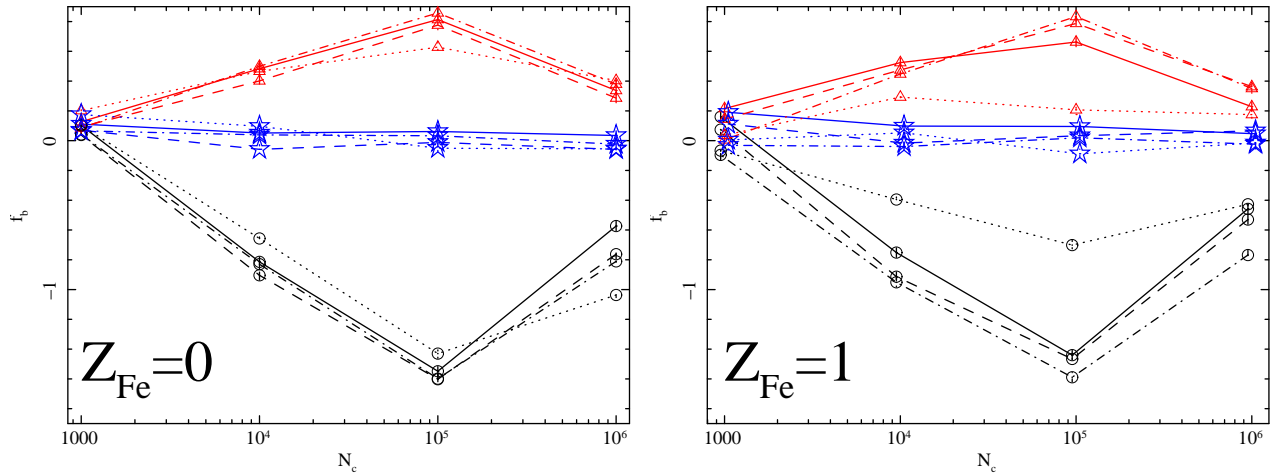}
\caption{\fb\ for the recovered temperature of a thermal plasma from its \chandra\ X-ray spectrum, 
as a function of total counts $N_c$ in the spectrum, for a thermal plasma with zero metal abundance
(\zfe$=0$) and for a Solar abundance plasma (\zfe$=1$). The data-points represent the results of 
the Monte Carlo simulations (see text), and the error-bars on \fb\ are all $\sim$0.03.
Circles denote \fb\ for the 
$\chi^2_d$ fits, triangles indicate $\chi^2_m$ fits, and stars indicate
C-statistic minimization. The temperature of the thermal plasma is indicated by the style
of the line joining the data-points, with solid, dashed, dot-dash and dotted lines
indicating a plasma with kT=1, 3, 5 and 7~keV, respectively.} \label{fig_spectrum}
\end{figure*}

We next consider the case of an X-ray emitting, 
optically thin, collisionally ionized thermal astrophysical plasma, the X-ray spectrum of which is
dominated by thermal bremsstrahlung plus line emission. Using the \xspec\
spectral-fitting package we simulated and fitted spectra which might be 
observed with the ACIS-I instrument aboard the \chandra\ X-ray observatory. Since 
we considered the high count limit, we did not include any background in
the simulations. For the source model we used a zero redshift APEC \citep{apec} plasma model modified
by line-of-sight absorption due to the cold Galactic ISM \citep{balucinska92}. We assumed an absorption
hydrogen column-density of $10^{20}$ $cm^{-2}$, consistent with 
a high Galactic latitude pointing.
The redistribution matrix (RMF) and effective area
(ARF) files (which map the physical source model onto the binned data taken by the detector) 
were created  for a near-aimpoint position in a representative 
ACIS-I observation. 

Using the ``fakeit'' command in \xspec\ we simulated sets of 1000 spectra 
for different combinations of temperature, metal abundance and 
total counts per spectrum. This procedure creates data in a set
of pre-defined bins by drawing a random number from a Poisson distribution with intrinsic 
mean $M_{i0}$, \ie\ the expected counts predicted by the model. 
We have verified that we obtain consistent
results with our own software. 
We chose input temperatures of 1, 3, 5 and 7 keV/k respectively and each model has heavy element
abundances relative to hydrogen set either to zero, or to match the Solar values
\citep{grsa}. We considered data only in the 0.5--7.0~keV range, and simulated
spectra with a range of $N_c$ spanning $10^3$ to $10^6$,
In each spectrum, the simulated data points were regrouped to ensure at least 20 photons per bin. 
We fitted each simulated spectrum while allowing only the temperature and normalization to vary,
separately using $\chi^2_d$, $\chi^2_m$ and the C-statistic, all of which are
implemented as standard in \xspec.
We show \fb\ obtained with each statistic as a function of temperature, heavy element abundance
and $N_c$ in Fig~\ref{fig_spectrum}.
In contrast to the $\chi^2_d$ and $\chi^2_m$ fits, for which $|$\fb$|\sim 0.5$--1 
it is immediately apparent that the C-statistic results are practially unbiased.

\section{Discussion} \label{sect_discussion}
For the two very different problems discussed in \S~\ref{sect_bias}, we find 
that, using realistic data, $|$\fb$|\ll1$ only for the fits using the C-statistic
 while the best-fitting parameters obtained by minimizing $\chi^2_d$ and $\chi^2_m$
were significantly biased (\fb\ of order unity). In order to explain these 
results, in the Appendices we derive an approximate analytical expression for the 
order of magnitude of 
\fb\ given an arbitrarily parameterized model. For fits using 
$\chi^2_d$, we find
\fb$\sim \mp N/\sqrt{N_c}$,
where N is the number of data-bins and $N_c$ the number of counts in the data-set. 
Alternatively, fits using $\chi^2_m$ were biased in the opposite sense, yielding
\fb$\sim \pm 0.5 N/\sqrt{N_c}$.
 This is true even in cases where the number of counts
far exceeds the canonical 20 per bin required for deviations from Gaussianity to
be unimportant. As pointed out by \citet{wheaton95a}, 
the bias arises not from deviations from Gaussianity but because of the 
misparameterization of the problem when these approximations are used with an arbitrary model. 
In contrast, those fits employing the C-statistic typically should have $|$\fb$|\ll1$. 
We show these
order of magnitude estimates for the lightcurve problem as the various lines 
in Fig~\ref{fig_lightcurve}, revealing excellent agreement with the results of our
simulations\footnote{In fact, for this problem, these estimates are almost exact, as can be 
seen by substituing the $M_{i}=p\tau$ into the derivations in the appendices.}.

The values of \fb\ obtained for the spectral-fitting problem (Fig~\ref{fig_spectrum}) 
are also easily understood
in terms of these order of magnitude estimates. 
In the regime of relatively few counts ($\sim$ 1000 per spectrum), the statistical errors can
be quite large (\eg\ $\pm3$~keV for the 7~keV plasma) and hence \fb\ was small for all
the statistics. For the cases 
with more counts the error-bars were small enough that the truncated Taylor expansion used in
the Appendices is approximately valid. 
Considering a typical 7~keV plasma with $N_c = 10^5$,
N is $\sim$400, implying  \fb$\sim-1.3$ for $\chi^2_d$ fits,
which is close to the observed value. As $N_c$ falls, so too does N since more data-bins need to 
be grouped together to ensure at least 20 counts in each. This can more than offset the fall in
$N_c$ and prevents \fb\ from growing much larger. In contrast, as $N_c$ gets 
even larger, there are few bins
at the original instrument resolution which contain fewer than 20 counts (\ie\ that 
need to be regrouped) and so $N$ grows only slightly
from $10^5$ to $10^6$ counts. Thus \fb\ starts to fall as $N_c$ gets very large, 
as seen in Fig~\ref{fig_spectrum}. A similar argument explains the trend of \fb\ with 
$N_c$ for the $\chi^2_m$ fits. 

\subsection{Removing the bias} \label{sect_removing_bias}
We have shown that, for fitting Poisson-distributed data with an arbitrary, parameterized
model {\em even in a fairly high-counts regime}, the routine use
of the $\chi^2_d$ and $\chi^2_m$ approximations to the true $\chi^2$ 
is likely to give rise to  biases in the best-fitting parameters which can be of order the 
statistical error, or even larger.
We argue, therefore, that the $\chi^2_d$ and $\chi^2_m$ approximations should
generally  be avoided for fitting Poisson-distributed data, unless the square root of the
number of counts in the dataset far exceeds the number of bins being fitted, or the 
model parameterization is chosen with care.
In contrast, fits performed using the \citeauthor{cash79a} C-statistic yield estimates which are,
to all practical purposes, unbiased in the regimes we have discussed in this paper and we,
therefore, strongly recommend its use instead.

The major objection to the 
widespread uptake of the C-statistic for model-fitting is that the statistic itself cannot be
directly interpreted as a goodness-of-fit indicator in a similar fashion to the (true) $\chi^2$
statistic.
In order to test the hypothesis that the data are consistent with the (best-fitting) model,
therefore one must adopt an alternative strategy. Arguably the most robust technique\footnote{For example,
the 
method outlined by \citet{baker84a} may not be accurate in all count regimes \citep{hauschild01a}.}
is a  fairly costly Monte Carlo approach, for example that implemented
as the ``goodness'' command in \xspec. On each simulation, an artificial
dataset is generated by adding Poisson-noise to the best-fitting model, and the artificial data
are fitted.
The fraction of simulations which yield a best-fitting statistic value which is more negative
(\ie\ a better fit) than the best-fit statistic for the real data is an estimate of 
the significance at which the null hypothesis can be rejected.
We note that the distribution of the best-fitting parameter values
from these simulations can be used at minimal extra computational cost to derive a confidence interval 
for each parameter
\citep[\eg][]{humphrey06a,buote03a}, as well as providing a direct assessment of the magnitude
of any residual bias. In the case where the number of fitted parameters becomes large, this Monte Carlo
method of error-bar estimation is far more efficient than the more usual procedure of 
stepping through parameter space \citep[\eg][]{cash79a}.

While it is not strictly necessary to bin the data in order to fit a model with the C-statistic,
the choice of binning is critical for interpreting the goodness-of-fit \citep[\eg][]{helsdon05a}.
The reason is that the statistic is defined only locally, in the sense that it contains no
information about the relative ordering of the residuals between data and model. To illustrate this
point, consider testing a lightcurve with the model $M_i=p\tau$.  Let the data be 
sufficiently sparsely binned that the number of counts in bin i, $D_i$ can only equal 0 or 1, 
and further let all of the
nonzero data-points be in the second half of the lightcurve (which clearly has only a 
$\sim 2^{-N_c}$ chance of occurring randomly, if the model is correct).
Substituting the best-fitting value ($p=N_c/N\tau$) into Eqn~\ref{eqn_cash}, it is clear that
\begin{eqnarray}
C& =  & 2\sum_i D_i log D_i -2 N_c log\left(\frac{N_c}{N}\right) = 
-2 N_c log\left(\frac{N_c}{N}\right) \nonumber
\end{eqnarray}
\ie\ C depends only on the number of counts in the lightcurve, and not their relative order. 
On each Monte Carlo simulation
we generate an artificial lightcurve from the best-fitting model, so clearly 
approximately half will have
more than $N_c$ counts in total, and half will have fewer. Provided $N_c/N \ll \exp(-1)$, which
must be true in this case, C varies monotonically with $N_c$ and so the estimated
null hypothesis probability will be 0.5 (\ie\ a ``good fit''). 
Alternatively, one can rebin the data into two equally-sized bins (one containing 0 counts and
one $N_c$), in which case the $\sum_i D_i log D_i$ term is no longer 0 and the test has greater
power to distinguish between the model and the data. Based on Monte Carlo simulations, 
 the model will be rejected at better than 99.9\%\ significance provided $N_c$\gtsim 8.
It is worth noting, however, that increasing the binning is not always helpful; if we 
were to bin the data even more heavily (into a single bin), we would wash out the information 
which allows us to distinguish between the model and the data. In the case that the 
data are inconsistent with the model, {\em the null hypothesis probability is almost always 
a strong function of the adopted binning}.

It is important to appreciate that the dependence of the null hypothesis probability on the binning
of the data is by no means limited to uses of the C-statistic, since $\chi^2$ 
(which also contains no information about the grouping of the residuals) suffers 
from exactly the same  problem \citep{gumbel43a}. In practice, the appropriate binning to use 
is that which maximizes the difference between the data and the model, which likely depends on the 
precise model being fitted and may involve some experimentation. Choosing to adopt the 
$\chi^2_m$ approximation on the grounds that it is ``easily interpretable'' for an ad hoc binning
scheme is clearly something of a false economy, especially coupled with the intrinsic bias which
can arise when it is used. 
The problem is exacerbated for the $\chi^2_d$ statistic, which is only
approximately $\chi^2$ distributed \citep{hauschild01a}.

Our present discussion does not consider the potential impact of background uncertainties 
\citep[which can introduce additional systematic errors; \eg][]{liu08a}, nor the case of very few 
counts per bin. In these circumstances it is possible that bias may remain on best-fitting 
parameters recovered from C-statistic fitting, or its variant in the \xspec\ package which 
takes account of direct background subtraction \citep{leccardi07a}. 
A full assessment of such putative effects needs to be carried out on a case-by-case basis, but
is relatively straightforward with the Monte Carlo method outlined above, and we will address
some of these issues in a future paper \citep{liu08a}.

Alternative approximations to $\chi^2$ have been proposed which are less biased 
in the case of  very few counts per bin (where the bias is partially due to
deviations from Gaussianity).
In general these schemes \citep[\eg][]{wheaton95a,kearns95a,churazov96a} 
are not rigorously motivated and there is no good theoretical reason to expect them to yield 
genuinely unbiased estimates for any given problem in the high counts case. 
Coupled with their lack of widespread use
and the difficulty of assessing their performance analytically, we do not address them here
other than to state that, aside from the ostensible transparency 
of the $\chi^2$ value (which, as stated above, can be deceptive), 
we see little compelling reason to use them in preference to the C-statistic.

\subsection{Scientific impact of the $\chi^2$ bias}
The existence of the bias will undoubtedly have implications for the scientific conclusions
of various studies which have adopted $\chi^2_d$ or $\chi^2_m$ approximations for fitting
Poisson distributed data without assessing the limitations of these approximations in that context.
In this section, we highlight a few cases of particular interest from the field
of X-ray astronomy, in which $\chi^2_d$ is typically adopted as a {\em de facto} 
standard (\eg\ in \xspec).

The in-flight inter-calibration of X-ray satellites can be assessed by comparing spectral-fits of
very bright, canonical ``calibration sources'' \citep[\eg][]{kirsch05a,plucinsky08a}.  Since different X-ray
instruments have different numbers of spectral bins (\eg\ typically \ltsim 500 for the \xmm\ PN 
and typically \ltsim 50 for the RossiXTE PCA) and since differences in exposure time and 
collecting area mean that there are widely varying numbers of photons in the calibration datasets,
the absolute magnitude  of the bias is expected to vary from instrument to instrument. 
For realistic sources it can be of order a few percent or higher, which
is  competetive with the absolute target calibration of most instruments. Since 
calibration sources are generally very bright, the statistical errors on recovered parameters are 
typically very small, and hence we may see parameter spaces which do not overlap even if the 
satellites are perfectly inter-calibrated. 

X-ray studies of galaxy clusters and groups routinely involve the computation of gravitating
mass profiles from the measured gas temperature and density profiles (obtained
from spatially-resolved spectroscopy) and the equation of hydrostatic equilibrium 
\citep[\eg][]{gastaldello07a}. Based on our 
simulations, and the arguments in Appendix~\ref{appendix_chi}, we expect roughly 
a 5--10\%\ fractional bias on the  temperature, which
would translate into a similar bias on the mass, especially in the cluster regime. 
Errors of this magnitude are significant if clusters
are to be used for precision cosmology measurements.
As an example, the relation between a cluster's virial mass (\mvir) and dark matter halo concentration (c), 
both of which are derived by fitting a canonical dark matter halo model (the NFW profile) to the 
measured mass profile, can be used  to distinguish between cosmological models.
Clearly \mvir\ is likely to be underestimated due to the bias
but the effect on c is harder to predict since it
depends sensitively on the exact slope of the mass profile, which in turn depends on how
the bias varies with radius. Still, if c is systematically biased by as much as $\sim$5\%,
as in our example below,
that would be comparable to the current best statistical error on the normalization of the 
c-\mvir\ relation,  which is the prime discriminator
between different cosmological models \citep{buote07a}.

To illustrate the bias on \mvir\ and c with real data, we have reduced and analysed high-quality \chandra\
data of a nearby, X-ray bright cluster, A\thin 1991. We obtained 39~ks of data from the \chandra\
archive, which we processed to obtain the temperature, gas density and gravitating mass
profiles as outlined in \citet{gastaldello07a}. Using the C-statistic we fitted the
data in 9 radial bins with parameterized models for the gas temperature and density which,
inserted into the equation of hydrostatic equilibrium, enabled us to 
obtain the mass profile and hence \mvir\ and c.
We found  \mvir$=2.60\pm 0.19 \times 10^{13}$\msun\ and c$=7.94\pm0.47$, which are broadly
consistent with the measurements of \citet{vikhlinin06b}, who apparently used $\chi^2_d$.
Refitting the data, this time using $\chi^2_d$, we found that the temperature was reduced by
$\sim$2\%\ on average and in individual bins it could change by as much as
$\sim$1-$\sigma$. This 
bias translated into a $\sim$4\%\ reduction in the resulting \mvir\ and c, or a
$\sim$0.5-$\sigma$ effect.
 The full details of this analysis will be given in \citet{liu08a}.

Another scaling relation which is key for understanding
cluster physics is the relation between \mvir\ and the emission-weighted X-ray temperature of the gas,
\tx. Both \tx\ and \mvir\ are likely underestimated in most published studies (which generally
use $\chi^2_d$). Since the spectrum used to measure the temperature usually contains far more 
counts than any of the individual spectra used to determine the mass profile, the effect on
\mvir\ is likely to be much larger. If this effect is as large as our estimated $\sim$5--10\%, 
it will not only  exceed the current best statistical error on the normalization of the measured 
relation, but it will also partially reduce the $\sim$30\%\ discrepancy in the normalization 
between the measured relation and the predictions
of self-similar models of cluster formation \citep[\eg][]{arnaud05a}.

As a final illustration of the effects of the bias in real data analysis, in his X-ray study of the
hot gas in galaxy groups, \citet{buote00c} estimated error-bars on the temperature 
and Fe abundance by a Monte Carlo
procedure similar to that discussed in \S~\ref{sect_removing_bias}. In a significant number of 
cases, the 1-$\sigma$ error range inferred from the simulations did not actually contain the 
best-fitting parameter (\ie\ the bias was more than 1-$\sigma$), giving rise to error-bars which
appeared distorted when plotted.

\acknowledgements
We thank H\'el\`ene Flohic for discussions.
Partial support for this work was provided by NASA under 
grant NNG04GE76G issued through the Office of Space Sciences Long-Term
Space Astrophysics Program. 
Partial support for this work was also provided by NASA through Chandra
Award Numbers G07-8083X and GO7-8131X issued by the Chandra X-ray
Observatory Center, which is operated by the Smithsonian Astrophysical
Observatory for and on behalf of NASA under contract NAS8-03060.
\appendix
\section{A. Bias in $\chi^2$-fitting} \label{appendix_chi}
\subsection{A1. Data weighting}
We here derive an expression for the magnitude of the bias when fitting a parameterized model
using the  $\chi^2_d$ approximation. 
We start by setting $M_i = M_i(p)$, with the parameter p having a 
true value $p_0$ and defining $M_{i0}=M_{i}(p_0)$. Starting with Eqn~\ref{eqn_chisq_data}, 
differentiating with respect to p and setting the derivative equal to zero, we obtain:
\begin{eqnarray}
0 & = & \frac{d \chi^2_d}{d p} = 2 \sum_i \frac{d M_i}{d p} \left( \frac{M_i-D_i}{D_i}\right)\nonumber \\
\end{eqnarray}
Now, we write $D_i=M_{i0}+\delta D_i$ and 
$M_i \simeq M_{i0}+\delta p M_{i0}^\prime+\delta p^2 M_{i0}^{\prime\prime}/2 +\ldots$, 
where $M_{i0}^\prime = dM_i/dp$ evaluated at $p=p_0$, and so on. Substituting these in and rearranging
we obtain:
\begin{eqnarray}
0 & = & \sum_i -\frac{M_{i0}^{\prime}}{M_{i0}} \delta D_i + \sum_i \frac{M_{i0}^{\prime}}{M_{i0}^2} \delta D_i^2  + 
   \delta p \left[ \sum_i \frac{M_{i0}^{\prime 2}}{M_{i0}}  + \sum_i \left( -\frac{M_{i0}^{\prime 2}}{M_{i0}^2} - \frac{M_{i0}^{\prime \prime}}{M_{i0}} \right)\delta D_i +
    \sum_i \left( \frac{M_{i0}^{\prime 2}}{M_{i0}^3} + \frac{M_{i0}^{\prime \prime}}{M_{i0}^2} \right) \delta D_i^2\right] \nonumber \\
 & & + \delta p^2 \left[ \sum_i \frac{3 M_{i0}^{\prime}M_{i0}^{\prime \prime}}{2 M_{i0}} +
	\sum_i \left(-\frac{3 M_{i0}^\prime M_{i0}^{\prime \prime}}{2 M_{i0}^2} -
	\frac{M_{i0}^{\prime \prime \prime}}{2 M_{i0}} \right) \delta D_i +
        \sum_i \left( \frac{3 M_{i0}^\prime M_{i0}^{\prime \prime}}{2 M_{i0}^3} +
	\frac{M_{i0}^{\prime \prime \prime}}{2 M_{i0}^2}\right) \delta D_i^2
 \right] + \ldots \nonumber \\
\end{eqnarray}
If higher order terms can be ignored, this is just a quadratic equation of the form:
\begin{eqnarray}
0 & = &  \sum_i a_i \delta D_i + \sum_i a_i^\prime \delta D_i^2 + 
\delta p \left(B+\sum_i b_i \delta D_i + \sum_i b_i^\prime \delta D_i^2\right) +\nonumber\\ & & 
\delta p^2 \left(C + \sum_i c_i \delta D_i  + \sum_i c_i^\prime \delta D_i^2 \right) \label{eqn_quadratic}\\
\Rightarrow \delta p & = &  \frac{-(B+\sum_i b_i \delta D_i+ \sum_i b_i^\prime \delta D_i^2)}
{2 (C+\sum_i c_i \delta D_i + \sum_i c_i^\prime \delta D_i^2)}
+\nonumber \\ & & \frac{\sqrt{(B+\sum_i b_i \delta D_i + \sum_i b_i^\prime \delta D_i^2)^2
-4 (\sum_i a_i \delta D_i + \sum_i a_i^\prime \delta D_i^2)(C + \sum_j c_j \delta D_j  + \sum_j c_j^\prime \delta D_j^2)}
}{2 (C+\sum_i c_i \delta D_i + \sum_i c_i^\prime \delta D_i^2)} \nonumber \\
\end{eqnarray}
where we only keep the solution consistent with \deltap\ being small. Assuming
$|\delta D_i| \ll M_{i0}$, both the square root and the recipricol terms can be expanded as a 
power series in $\delta D_i$. Writing only terms up to second order, we obtain:
\begin{eqnarray}
\delta p & \simeq & -\frac{1}{B}\sum_i a_i \delta D_i -\frac{1}{B}\sum_i a_i^\prime \delta D_i^2 +
\sum_{ij} \frac{\delta D_i \delta D_j}{B^2} \left(\frac{1}{2}(b_i a_j + b_j a_i) - \frac{C a_i a_j}{B}\right) \label{eqn_deltap}\nonumber \\ 
\Rightarrow <\delta p> & \simeq & \sum_i \frac{M_{i0}}{B^2} \left( b_i a_i - \frac{C a_i^2}{B} - B a_{i}^\prime \right) \label{eqn_bias_data}
\end{eqnarray}
where $<$\ldots$>$ denotes the expectation operator. 
We have used the distributive nature of the expectation operator and we have used the results
$<\delta D_i>\equiv 0$ and $<\delta D_i \delta D_j> \equiv 0$, if $i\ne j$
or $=M_{i0}$ if $i=j$, which are true 
for both Poisson and Gaussian distributions (provided the latter has a statistical error in 
bin i, $\sigma_i=\sqrt{<D_i>}$). 

In general, $<\delta p>$ will be nonzero. To estimate its magnitude it is helpful
to define $M_{i0}^\prime \equiv M_{i0} f_i^\prime(p_0)/p_0$,  
$M_{i0}^{\prime\prime} \equiv M_{i0} f_i^{\prime \prime}(p_0)/p_0^2$ and $M_{i0}\equiv N_c m_{i0}$,
 where $N_c$ is the total number of counts in the dataset.
Making these substitutions and rearranging we find that 
\begin{eqnarray}
b_i a_i & = & \frac{f_i^{\prime}}{p_0^3}(f_i^{\prime 2}+f_i^{\prime \prime}),\hspace{0.5cm} 
-\frac{C}{B}a_i^2 = -\frac{3 f_i^{\prime 2} \overline{f^{\prime} f^{\prime \prime}}}{2 p_0^3 \overline{f^{\prime 2}}}\hspace{0.5cm} {\rm and}
-Ba_i^\prime = -\frac{f_i^\prime \overline{f^{\prime 2}}}{p_0^3 m_{i0}} \nonumber \\
\end{eqnarray}
where $\overline{f^{\prime 2}}\equiv \sum_i f_i^{\prime 2}m_{i0}$, \ie\ the model-weighted
average of $f_i^{\prime 2}$, and so on. We note that
\begin{eqnarray}
\frac{M_{i0} f_i^{\prime \prime}}{p_0^2} & = & M_{i0}^{\prime\prime} = \frac{d M_{i0}^\prime}{d p_0} =
\frac{M_{i0}}{p_0^2} \left(f_i^{\prime 2}-f_i^\prime +p_0\frac{d f_i^\prime}{dp_0}\right) \Rightarrow
 f_i^{\prime \prime} = f_i^{\prime 2}-f_i^\prime +p_0\frac{d f_i^\prime}{dp_0} \label{eqn_fiprime}
\end{eqnarray}
and so, on average, ${f_i^{\prime} f_i^{\prime \prime}} \sim {f_i^{\prime 3}}$ for a broad class 
of problem, where the $\sim$ symbol indicates similar orders of magnitude. Thus, on average $b_i a_i \sim f_i^{\prime 3}/p_0^3$, 
$-Ca_i^2/B \sim -f_i^{\prime 3}/p_0^3$ and $-B a_i^\prime \sim -N f_i^{\prime 3}/p_0^3$, where we have used
$1/m_{i0}\sim N$, the number of data bins. Since $N\gg 1$ in general, it follows that the third of the parenthetical 
terms in Eqn~\ref{eqn_bias_data} is much larger than the other two. Keeping only that term, Eqn~\ref{eqn_bias_data}
becomes
\begin{eqnarray}
<\delta p> \simeq -\frac{p_0 N}{N_c}\left[ \frac{\sum_i \frac{1}{N} f_i^\prime}{\overline{f^{\prime 2}}}\right] \nonumber \\
\end{eqnarray}

To estimate \fb,  we adopt the statistical error obtained from fitting 
the C-statistic, which is expected to be close to that obtained with $\chi^2$ methods
\citep{cash79a}.
As we show in Appendix~\ref{appendix_cash}, to second order this is given
by:
\begin{eqnarray}
<\delta p^2> \simeq \frac{\sum_i M_{i0}^{\prime 2} M_{i0}}{\left(\sum_i M_{i0}^{\prime 2}\right)^2}
=\frac{p_0^2}{N_c} \frac{\sum_i m_{i0}^3 f_i^{\prime 2}}{\left(\sum_i m_{i0}^2 f_i^{\prime 2}\right)^2}
\sim \frac{p_0^2}{N_c \overline{f^{\prime 2}}} \label{eqn_error}
\end{eqnarray}
We have assumed 
$\sum_i m_{i0}^{j} f_i^{\prime 2} \sim \overline{f^{\prime 2}}/N^{j-1}$, which is justified since
$m_{i0}\sim 1/N$. Thus we obtain:
\begin{eqnarray}
f_b & \sim&  - \frac{N}{\sqrt{N_c}}\left[\frac{\sum_i \frac{1}{N} f_i^\prime}{\sqrt{\overline{f^{\prime 2}}}} \right]
\sim \mp \frac{N}{\sqrt{N_c}} \label{eqn_fb_data}
\end{eqnarray}
where we have assumed the term in square brackets is $\sim \pm 1$, that is the absolute value of the 
mean of $f_i^\prime$ (averaged over the data set) is of the same order of magnitude as its 
(model-weighted) root mean square. This will likely be approximately true for an arbitrary
model (although it should be verified in any particular case) unless one takes considerable
care over choosing the particular parameterization of the model, in which case it may
be possible to obtain \fb\ close to zero. 

Strictly speaking, this derivation is only valid for single-parameter models. However, it is 
relatively straightforward to generalize it to the multi-parameter case, which leads to 
 a set of coupled quadratic equations (one per parameter) of a form similar to 
Eqn~\ref{eqn_quadratic}. This implies that the bias on the parameters, or at least some
combination of the parameters, should be of a similar order to that derived above.

\subsection{A2. Model weighting}
For the case of model weighting, the problem is remarkably similar. Starting with 
Eqn~\ref{eqn_chisq_model} differentiating and rearranging, we obtain
\begin{eqnarray}
0 & = & \frac{d \chi^2_m}{d p} = \sum_i \frac{d M_i}{d p} \left( \frac{M_i^2-D_i^2}{M_i^2}\right)\nonumber\\
\end{eqnarray}
Using the same expansion methods we adopted for the data-weighting case, we obtain (ignoring all terms
higher than second order):
\begin{eqnarray}
0 & \simeq & \sum_i -\frac{2 M_{i0}^\prime}{M_{i0}} \delta D_i +  \sum_i -\frac{M_{i0}^\prime}{M_{i0}^2}\delta D_i^2 + \delta p \left[\sum_i \frac{2 M_{i0}^{\prime 2}}{M_{i0}} + \sum_i \left(\frac{4 M_{i0}^{\prime 2}}{M_{i0}^2} - \frac{2 M_{i0}^{\prime \prime}}{M_{i0}} \right)\delta D_i + \right.\nonumber\\
& & \left. \sum_i\left(\frac{2 M_{i0}^{\prime 2}}{M_{i0}^3} - \frac{M_{i0}^{\prime \prime}}{M_{i0}^2} \right) \delta D_i^2 \right]
+ \delta p^2 \left[\sum_i \left(\frac{3 M_{i0}^{\prime}M_{i0}^{\prime \prime}}{M_{i0}} -
\frac{3 M_{i0}^{\prime 3}}{M_{i0}^2} \right) \right. \nonumber\\ & & + \left.\sum_i \left( -\frac{6 M_{i0}^{\prime 3}}{M_{i0}^3} +
\frac{6 M_{i0}^\prime M_{i0}^{\prime \prime}}{M_{i0}^2} - \frac{M_{i0}^{\prime \prime\prime}}{M_{i0}} \right) \delta D_i  +\sum_i \left(-\frac{3 M_{i0}^{\prime 3}}{M_{i0}^4} + \frac{3 M_{i0}^\prime M_{i0}^{\prime \prime}}{M_{i0}^3} - \frac{M_{i0}^{\prime\prime\prime}}{2 M_{i0}^2}\right) \delta D_i^2 \right] \nonumber \\
\end{eqnarray}
which is a quadratic in $\delta p$, of the form discussed in the previous section. Therefore, the bias
can be trivially computed from Eqn~\ref{eqn_bias_data}. Substituting for $M_{i0}$, $M_{i0}^\prime$ and 
$M_{i0}^{\prime \prime}$, exactly as before, we obtain
\begin{eqnarray}
b_ia_i & \sim & \frac{4 f_i^\prime}{p_0^3}\left(f_i^{\prime \prime}-2 f_i^{\prime 2}\right),\hspace{0.5cm}
  -\frac{C a_i^2}{B} \sim \frac{6 f_i^{\prime 2}}{p_0^3} \left( \frac{\overline{f^{\prime 3}}-\overline{f^\prime f^{\prime \prime}}}{\overline{f^{\prime 2}}}\right)
\hspace{0.5cm} {\rm and}\ -B a_i^\prime \sim  \frac{2 f_i^\prime \overline{f^{\prime 2}}}{m_{i0}p_0^3}  \nonumber \\
\end{eqnarray}
Following the arguments used for the data-weighting case, it is clear that $|B a_i^\prime|$ is much 
larger than the other terms, so 
\begin{eqnarray}
 <\delta p> & \simeq & \frac{1}{2} \frac{p_0 N}{N_c} \left[ \frac{\sum_i \frac{1}{N} f_i^\prime}{\overline{f^{\prime 2}}}\right]
\Rightarrow f_b \sim \frac{1}{2}\frac{N}{\sqrt{N_c}}\left[\frac{\sum_i \frac{1}{N} f_i^\prime}{\sqrt{\overline{f^{\prime 2}}}} \right]
\sim \pm \frac{1}{2}\frac{N}{\sqrt{N_c}} 
\end{eqnarray}
Note that the bias due on parameters recovered under the $\chi^2_d$ approximation
is $-2$ times the bias with $\chi^2_m$.

\section{B. Cash C-statistic bias and error} \label{appendix_cash}
We here estimate the magnitude of the bias and the 
statistical error we expect on the recovered parameter for
the case where the \citeauthor{cash79a} C-statistic is used to fit the data. In general, it is 
expected that parameters obtained from a maximum likelihood method have some level of bias
\citep[\eg][]{ferguson82a} but we here show that, for the C-statistic in the high counts regime,
this bias is likely far smaller than the statistical error.
We can approach
this problem by essentially the same technique used in Appendix~\ref{appendix_chi}.
Differentiating Eq~\ref{eqn_cash}, setting it equal to 0 and 
rearranging, we obtain:
\begin{eqnarray}
0&=&\sum_i \frac{d M_i}{d p} \left(\frac{M_i-D_i}{M_i}\right) \nonumber \\
\end{eqnarray}
Using the expansion methods we adopted in Appendix~\ref{appendix_chi}, we obtain the approximate expression:
\begin{eqnarray}
0& =& -\sum_i M_{i0}^\prime \delta D_i +  \delta p \left[ \sum_i M_{i0}^{\prime 2} + \sum_i \left( -M_{i0}^{\prime \prime} + 
\frac{M_{i0}^{\prime 2}}{M_{i0}}\right)\delta D_i  \right] \nonumber \\
& & + \delta p^2 \left[ \left( \sum_i \frac{3}{2} M_{io}^{\prime \prime}M_{io}^\prime - 
     \frac{M_{i0}^{\prime 3}}{M_{i0}} \right)  
    + \sum_i \left( -\frac{M_{i0}^{\prime \prime \prime}}{2}+\frac{M_{i0}^\prime M_{i0}^{\prime \prime}}{2 M_{i0}} - \frac{M_{i0}^{\prime 3}}{M_{i0}^2}\right) \delta D_i \right] + \ldots
\end{eqnarray}
If higher order terms can be ignored, this is just a quadratic equation similar to that
solved in Appendix~\ref{appendix_chi}, but with $a_i^\prime=b_i^\prime=c_i^\prime= 0$. 
From Eqn~\ref{eqn_deltap} it is easy to 
show that only keeping terms up to second order,
\begin{eqnarray}
<\delta p^2> & \simeq & \frac{1}{B^2} \sum_{ij} a_i a_j <\delta D_i \delta D_j> =
\frac{1}{B^2} \sum_{i} a_i^2 M_{i0} = \frac{\sum_i M_{i0}^{\prime 2} M_{i0}}{\left( \sum_i M_{i0}^{\prime 2} \right)^2}\label{eqn_staterr}
\end{eqnarray}

Now, in general the C-statistic fits are found to be far less biased than those using 
$\chi^2_d$ or $\chi^2_m$. This can be shown by 
substituting the appropriate expressions for each of the terms in Eqn~\ref{eqn_bias_data}
and making the various substitutions for
$M_{i0}$, $M_{i0}^\prime$ and $M_{i0}^{\prime\prime}$ outlined in Appendix~\ref{appendix_chi}. We 
obtain:
\begin{eqnarray}
<\delta p> & \simeq& \frac{p_0}{N_c} \left[\frac{\sum_j m_{j0}^2 f_j^{\prime 2}\sum_i\left(m_{i0}^3 f_i^\prime f_i^{\prime\prime}- f_i^{\prime 3} m_{i0}^3\right)
-\sum_j m_{j0}^3f_j^{\prime 2} \sum_i \left(\frac{3}{2} m_{i0}^2 f_i^\prime f_i^{\prime\prime}- f_i^{\prime 3} m_{i0}^2 \right)
}{\left(\sum_j m_{j0}^2 f_j^{\prime 2}\right)^3}
\right] \nonumber \\
\end{eqnarray}
Now, assuming $\sum_i m_{i0}^{k} f_i^{\prime 2} \sim \overline{f^{\prime 2}}/N^{k-1}$ (see Appendix~\ref{appendix_chi}),
we obtain:
\begin{eqnarray}
<\delta p> & \sim & \frac{p_0}{N_c} \left[ -\frac{\overline{f^\prime f^{\prime \prime}}}{2 \left(\overline{f^{\prime 2}}\right)^2}\right] \nonumber \\
\end{eqnarray}
where we have allowed two terms of order $\overline{f^{\prime 3}}$ in the numerator of the bracketed expression to cancel; 
although they are unlikely
to cancel completely we assume that they largely do so, making the
$\overline{f^\prime f^{\prime\prime}}$ term more important. Relaxing this assumption does not affect our
conclusions. Adopting the order of magnitude estimate for the statistical error derived in Appendix~\ref{appendix_chi},
we obtain 
\begin{eqnarray}
f_b \sim \frac{1}{\sqrt{N_c}}\left[ -\frac{\overline{f^\prime f^{\prime \prime}}}{2 \left(\overline{f^{\prime 2}}\right)^\frac{3}{2}}\right] \sim \mp \frac{1}{\sqrt{N_c}} \label{eqn_fb_cash}
\end{eqnarray}
which is vanishingly small as $N_c$ becomes large. We have assumed that the term in square brackets is of order
unity. This can be justified because, as shown in Appendix~\ref{appendix_chi},
$\overline{f^{\prime} f^{\prime \prime}}\sim \overline{f^{\prime 3}}$ which $\sim (\overline{f^{\prime 2}})^{3/2}$
for a broad range of problem. Although the accuracy of this assumption should be tested for any given problem,
provided $\overline{f^{\prime} f^{\prime \prime}}$ is not larger than $(\overline{f^{\prime 2}})^{3/2}$
by a factor $\sim N (\gg 1)$, the parameters recovered from the C-statistic fit will be less biased than
those using $\chi^2_d$ or $\chi^2_m$. Finally, since typically $f_b \ll 1$ we are justified in
assuming  $\sqrt{<\delta p^2>}$ is the 1-$\sigma$ statistical error on p.

\bibliographystyle{apj_hyper}
\bibliography{paper_bibliography.bib}

\end{document}